\documentclass[onecolumn,secnumarabic,amssymb, nobibnotes, aps, prd]{revtex4}
\usepackage{graphicx}

\begin{document}
\title{A development of superconducting differential double contour interferometer}
\author{V.L. Gurtovoi$^{1,2}$, V.N. Antonov$^{2,3}$, A.V. Nikulov$^{1}$, R. Shaikhaidarov$^{3}$, and V.A. Tulin$^{1}$}
\affiliation{$^{1}$Institute of Microelectronics Technology and High Purity Materials, Russian Academy of Sciences, 142432 Chernogolovka, Moscow District, Russia. \\ $^{2}$Moscow Institute of Physics and Technology, 29 Institutskiy per., 141700 Dolgoprudny, Moscow Region, Russia. \\ $^{3}$Physics Department, Royal Holloway University of London, Egham, Surrey TW20 0EX, UK } 
\begin{abstract} We study operation of a new device, the superconducting differential double contour interferometer (DDCI), in application for the ultra sensitive detection of magnetic flux and for digital read out of the state of the superconducting flux qubit. DDCI consists of two superconducting contours weakly coupled by Josephson Junctions. In such a device a change of the critical current and the voltage happens in a step-like manner when the angular momentum quantum number changes in one of the two contours. The DDCI may outperform traditional Superconducting Quantum Interference Devices when the change of the quantum number occurs in a narrow magnetic field region near the half of the flux quantum due to thermal fluctuations, quantum fluctuations, or the switching a loop segment in the normal state for a while by short pulse of an external current. 
\end{abstract}

\maketitle  


It is about 50 years since first measurements of Superconducting Quantum Interference Device (SQUID) \cite{SQUID1964}, which is one of the most sensitive detectors of magnetic field \cite{SQUID1977}. The classical dc SQUID consists of two Josephson junctions mounted in a superconducting loop with a square $S$. The critical current $I _{c}$ of the two junctions oscillates as a function of the external magnetic flux $\Phi = BS$ threading the loop, with the period equal the flux quantum $\Phi _{0} = 2\pi \hbar /2e \approx 20.7  \ Oe \ \mu m^{2}$ \cite{SQUID1977}. The voltage $V$ across the junctions at constant bias current $I$ is also periodic in $\Phi $. The classical dc SQUID is used as the detector of magnetic flux $\Phi $ or magnetic field $B = \Phi /S$ due to the periodic dependence $V(\Phi )$ at $I = const $. Therefore the maximum value $(\partial V/\partial \Phi ) _{I}$ is important parameter for the dc SQUID sensitivity \cite{SQUID1977}. The amplitude of the voltage oscillations $\Delta V = R _{d}\Delta I _{c}$ cannot exceed the value $R _{d}I _{c} < \Delta /e$, where $ R _{d}$ is the dynamical resistance of the Josephson junctions, $\Delta $ is the energy gap of the superconductor and $e$ is the electron charge \cite{Barone}. According to the relation $I _{c} = 2 I _{c,j}|\cos \pi \Phi |$ \cite{Barone} valid in the case of week screening $\beta = 2L I _{c,j}/\Phi _{0} \ll 1$  \cite{SQUID1977} the critical current of the dc SQUID changes in the interval $\Delta \Phi = \Phi _{0}/2$. Consequently the maximum value $(\partial V/\partial \Phi ) _{I}$ of the classical dc SQUID cannot exceed $2\Delta /e\Phi _{0}$. The real value $(V/\Phi ) _{I} \approx 2 \ \mu V/\Phi _{0}$ \cite{SQUID1977} of a typical dc SQUID is substantially smaller than the maximum value. In this work we explore an alternative device for the measurement of the weak magnetic filed, a differential double contour interferometer (DDCI). The idea of this new device arose thanks to the experiment made in \cite{Zhilyaev2000} and its explanation \cite{Zhilyaev2001}.

Higher sensitivity of DDCI compared to the conventional SQUID is provided by strong discreteness of the energy spectrum of the continuous superconducting loop \cite{PLA2012QF}. According to the conventional theory \cite{Tink75} the total energy of the persistent current 
$$I_{p} = \frac{n\Phi_{0} - \Phi }{L_{k}} \eqno{(1)}$$
in a loop with small cross
section $s \ll \lambda _{L}^{2}(T)$ is determined mainly by the kinetic energy \cite{QuSMF2016}:  
$$E _{t} \approx E _{k} = \frac{L_{k}I_{p}^{2}}{2}= \frac{\Phi _{0}^{2}}{2L_{k}}(n - \frac{\Phi }{\Phi _{0}})^{2} \eqno {(2)}$$
Here $L _{k} = ml/ q^{2}n_{s}s = (\lambda _{L}^{2}/s) \mu _{0}l \approx (\lambda _{L}^{2}/s)L$ is the kinetic inductance of the loop of side $l$; $L \approx \mu _{0}l$ is the magnetic inductance; $s$ is the cross section of superconducting wires; $n _{s}$ is the density of the Cooper pairs; $\lambda _{L} = (m/\mu _{0}q^{2}n_{s})^{0.5} $ is the London penetration depth. 

Two permitted states $n$ and $n+1$ have equal energy at $\Phi = (n+0.5) \Phi _{0} $ according to (2) and thus equal probability $P \propto \exp (-E _{k}/k _{B}T) $. The probability of other permitted states is negligible and $P(n+1) \approx 1 - P(n)$ at $\Phi \approx (n+0.5) \Phi _{0} $ when $\Phi _{0}^{2}/2L_{k} = I _{p,A}\Phi _{0} \gg k _{B}T$. The probability of the $n$ state may be described with the relation  
$$P(n)  \approx  \frac{1}{1+\exp \epsilon 2\frac{\delta \Phi}{\Phi _{0}}} \eqno {(3)}$$ 
at the magnetic flux $\Phi = (n+0.5) \Phi _{0} + \delta \Phi $, when $\delta \Phi \ll \Phi _{0}$. Here $ I _{p,A} = \Phi _{0}/2L_{k}$ is the persistent current (1) at $|n - \Phi /\Phi _{0}| = 1/2$ and $\epsilon = I _{p,A}\Phi _{0}/k _{B}T $. The probability (3) changes from $P(n) \approx 1$ to $P(n) \approx 0$ in a region of the magnetic flux from $\delta \Phi \approx  -\Phi _{0}/2\epsilon $ to $\delta \Phi \approx  \Phi _{0}/2\epsilon $. This region may be very narrow due to a big value $\epsilon = I _{p,A}\Phi _{0}/k _{B}T \gg 1$ equal, for example $\epsilon \approx 1500$ at the temperature of measurement $T  \approx 1 \ K$ and a typical value $I _{p,A} = 10 \ \mu A$ measured, for example in \cite{PCJETP07}. Measurements \cite{Tanaka04} of flux qubit (superconducting loop with three Josephson junctions) corroborate the change of the probability $P(n) $ in a narrow region of magnetic flux predicted by the relation (3). For example the probability changes from $P(n) \approx 1$ to $P(n) \approx 0$ in a region from $\delta \Phi \approx  -0.002\Phi _{0}$ to $\delta \Phi \approx  0.002\Phi _{0}$ at the temperature of measurement $T  \approx 0.1 \ K$, see Fig.5 of \cite{Tanaka04}, and the value $I _{p,A} < 0.6 \mu A$  of the flux qubit measured in \cite{Tanaka04}.

\begin{figure}[]
\includegraphics{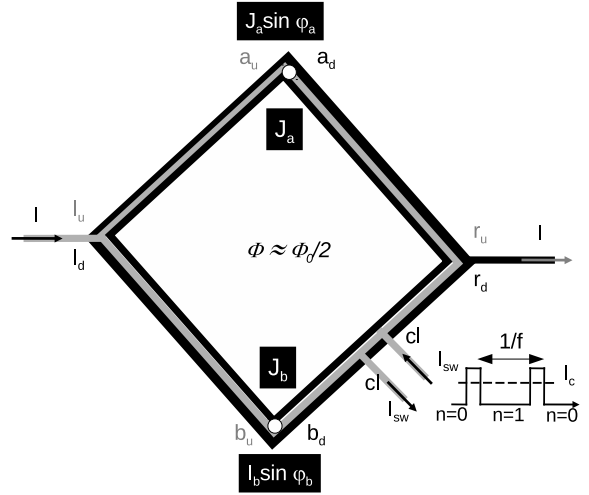}
\caption{\label{fig:epsart} The work principle of the superconducting differential double contour interferometer (DDCI). A bias current $I$ flows from the upper loop to the lower loop through two Josephson junctions $J _{a}$ and $J _{b}$  (labeled with white circles). The maximum value of the superconducting current through these Josephson junctions should depend on parity of the sum $n _{u} + n _{d}$ of the quantum numbers of the upper loop $n _{u}$ and the lower loop $n _{d}$ according to (6). Therefore the voltage measured on DDCI average in the time $\overline{V} = \Theta ^{-1}\int _{\Theta} dt V(t) \approx V _{min}P(n _{u} =0) + V _{max}P(n _{u} =1)$ should change from the minimum value $\overline{V} \approx V _{min}$ to the maximum value $\overline{V} \approx V_{max}$ in the narrow interval $\Delta \Phi \approx \Phi _{0}/\epsilon \ll \Phi _{0}/2$ at $\epsilon \gg 1$ when $n _{d} = 0$ and the quantum number of the upper loop takes two values $n _{u} = 0$ and $n _{u} = 1$ due to the switching of a segment of this loop in normal state by the short pulses of an external current $I_{sw}$ with the amplitude exceeding the critical current $I_{c}$ of the superconducting strip.}
\end{figure} 

The quantum number of the continuous superconducting loop, in contrast to the flux qubit \cite{Tanaka04}, can change only if it or its segment is switched in normal state for a while \cite{PRB2014C}. The numerous measurements of the critical current \cite{PCJETP07,APL2016,JETP07J} and other parameters \cite{PCScien07,PLA2012} testify that the loop comes back in the superconducting state with the quantum number $n$ corresponding to the minimal kinetic energy (2) at the  $(n-0.5)\Phi _{0} < \Phi < (n+0.5) \Phi _{0}$ with the predominant probability $P(n) \approx 1$.The probability $P(n)$ of the $n$ state should change in a narrow interval of magnetic flux $\Delta \Phi \approx  \Phi _{0}/\epsilon $ near $\Phi = (n+0.5) \Phi _{0} $ when a segment of the continuous superconducting loop is switched in the normal state by the short pulses of an external current $I_{sw}$ with a frequency $f$. Such switching may be provided with the help of the additional current leads shown on Fig.1. The quantum number $\overline{n} = \Theta ^{-1}\int _{\Theta} dt n(t)$ average in a time $\Theta \gg 1/f$  equal $\overline{n} \approx nP(n) + (n+1)P(n+1)$ should change in this narrow interval $\Delta \Phi \approx \Phi _{0}/\epsilon \ll \Phi _{0}/2$ from $n$ to $n+1$. We propose to use the differential double contour interferometer, shown on Fig.1, in order to transform the $\overline{n}$ variation in the variation of the voltage average in the time $\overline{V} = \Theta ^{-1}\int _{\Theta} dt V(t)$ in the same narrow interval $\Delta \Phi \ll \Phi _{0}/2$. It may be possible due to the jump of the critical current of the DDCI with the change of the quantum number of one of its contours from $n$ to $n+1$ predicted in \cite{NANO2010}. The superconducting current between points $l _{u}$ and  $r _{d}$ of DDCI, Fig.1, equals the sum 
$$I _{s} = I _{a}\sin  \varphi _{a} + I _{b}\sin  \varphi _{b} \eqno{(4)}$$ 
of the currents $I _{a}\sin  \varphi _{a}$ and $I _{b}\sin  \varphi _{b}$ through the Josephson junctions $J _{a}$ and $J _{b}$. Here $I _{a}$ and $I _{b}$ are the critical currents of the Josephson junctions; $\varphi _{a}$ and $\varphi _{b}$ are the phase differences between the up $a _{u}$, $b _{u}$ and down $a _{d}$, $b _{d}$ boundaries of the Josephson junctions. The relation 
$$\oint_{l}dl \bigtriangledown \varphi = 2\pi n \eqno{(5)}$$
must be valid for the both contours $l _{u} - a _{u}- r_{u} - b_{u}- l _{u}$ and $l _{d} - a _{d}- r_{d} - b_{d}- l _{d}$ due to the requirement that the complex wave function must be single-valued in any point of the circumference $l $ of each contour $\Psi = |\Psi |e^{i\varphi } =  |\Psi |e^{i(\varphi + n2\pi )}$. The relation (5) should be valid also for the contours $l _{u} - a _{u} - a _{d} - r_{d} - b_{d}- b_{u}- l _{u}$ when the current through the Josephson junctions does not exceed the critical current $I _{a}$ and $I _{b}$. Then, the superconducting current between $l _{u}$ and  $r _{d}$ is equal to:  
$$I _{s} = I _{a}\sin  \varphi _{a} + I _{b}\sin  (\varphi _{a}+ \pi (n _{u} + n _{d})) \eqno{(6)}$$
The critical current $I _{c,DD}$ of the DDCI, i.e. the maximum value of the superconducting current (6), depends only on parity of the quantum number sum, and it does not explicitly depend on the magnetic flux, which makes the system to be an ideal detector of the quantum states. The critical current has only two values at $I _{a}= I _{b} = I _{cj}$ according to (6): $I _{c,DD} = 2 I _{c,j}$ when the sum $n _{u} + n _{d}$ of the quantum numbers of the upper loop $n _{u}$ and the bottom loop $n _{d}$ is even and $I _{c} = 0$ when it is odd. Therefore, for example at $\Phi = 0.5 \Phi _{0} + \delta \Phi $, the maximum value of the superconducting current (6) and the voltage measured on the DDCI at a bias current $I$ may change by jump when the quantum number of the  bottom loop should be equal permanently zero $n _{d} = 0$ whereas the quantum number the upper loop may change from $n _{u} =0$ to $n _{u} =1$ (or from $n _{u} =1$ to $n _{u} = 0$) after successive switching of the loop segment in the normal state by the short pulse of the external current $I_{sw}$, Fig.1. The probability of the minimum value of the voltage $V _{min}$ (the maximum value of the critical current $I _{c,DD} = 2 I _{c,j}$ ) should be equal the probability (3) of the state $n _{u} =0$ because $n _{u} + n _{d} = 0 + 0 = 0$ is even. Therefore the voltage average in the time $\overline{V} = \Theta ^{-1}\int _{\Theta} dt V(t) \approx V _{min}P(n _{u} =0) + V _{max}P(n _{u} =1)$ should change from the minimum value $\overline{V} \approx V _{min}$ to the maximum value $\overline{V} \approx V_{max}$ in the narrow interval $\Delta \Phi \approx \Phi _{0}/\epsilon \ll \Phi _{0}/2$ at $\epsilon \gg 1$. The DDCI may be more sensitive than the conventional dc SQUID due to this sharpness of the voltage $\overline{V}$ change at its most sensitive point $\Phi \approx 0.5 \Phi _{0} $ (or $\Phi \approx (n+0.5) \Phi _{0} $). One may expect to increase the flux sensitivities in $\epsilon /2$ times if the jump $V_{max} - V _{min}$ is not less than the voltage variation of the conventional dc SQUID.  


In order to ensure that the change of the quantum number $n _{u}$ results to the voltage jump and to measure its value we used a real structures shown on Fig.2. The structure was fabricated by e-beam lithography to form suspended resist mask and double angle shadow evaporation of aluminum ($d \approx 30 \ nm$ and $d \approx 35 \ nm$) with intermediate first aluminum layer oxidation. This technology allows to make two independent superconducting square contours weakly connected by two Josephson junctions in the two points $J _{a}$ and $J _{b}$, Fig. 1. The contours are shifted relative to one another and there are two extra Josephson junctions $J _{t1}$ and $J _{t2}$ because of this technology. The structures with square side of the loops $a \approx  4 \ \mu m$ and $a \approx  20 \ \mu m$ were made and investigated. The width of the loops was $w \approx 400 \ nm$ and their cross section $s = dw \approx 12000 \ nm^{2}$ corresponds to $\lambda _{L}^{2}(T) = \lambda _{L}^{2}(0)(1 - T/ T _{c})^{-1}$ of aluminium with its London penetration depth $\lambda _{L}(0) \approx 50 \ nm$ at $T \approx 0.8T _{c}$. The critical temperature of aluminium loops was $T _{c} \approx 1.3 \ K$. 

\begin{figure}[]
\includegraphics{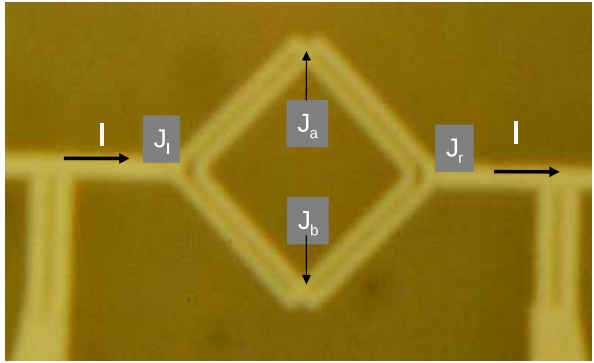}  
\caption{\label{fig:epsart} The real structure used for the observation of the voltage jumps with the change of the quantum numbers of its two loops weakly connected by two Josephson junctions $J _{a}$ and $J _{b}$. The two aluminium loops are shifted with respect to each other because of the simple technology of two-angle evaporation used in our work. The structure has two unnecessary Josephson junctions $J _{l}$ and $J _{r}$ because of this technology.  Photo of 4 $\mu m $ aluminium loops is shown.}  
\end{figure}

The structure, shown on Fig.2, without the additional current leads $cl$, Fig.1, cannot be used as a sensitive magnetometer because the quantum number $n _{u}$ of the continuous superconducting loop may not meet the minimum kinetic energy (2) when no segment is switched in the normal state. We use this structure only for the observation of the voltage jump due to the change of the quantum number. This change takes place when the persistent current (1) increases in the loop without the current leads with the change of the magnetic flux $\Phi = BS + LI_{p}$ up to a critical value \cite{Geim2003,Moler2007,Nat2016}. The pair velocity $v = (2\pi \hbar/ml)(n - \Phi /\Phi _{0})$  can increase with the $\Phi $ variation up to the depairing velocity $v _{c} = \hbar /m \surd 3 \xi (T)$ \cite{Tink75} but cannot exceed this value $|v| \leq v _{c}$. Therefore the quantum number should change at $|n - \Phi /\Phi _{0}| \leq l/2\pi \surd 3 \xi (T)$ \cite{Geim2003,Moler2007,Nat2016}. The magnetic flux at which the quantum number $n$ changes may be not equal $\Phi = (n+0.5)\Phi _{0}$ and may be different in ascending and descending magnetic field.

The jumps of the critical current (6) and the voltage at a non-zero bias current $I \neq 0$ can be observed only if the quantum numbers of the upper loop $n _{u}$ and the lower loop $n _{d}$ change at different $\Phi $ values. The voltage may jump up (when the sum $n _{u} + n _{d}$ becomes an odd number) and down (when the sum $n _{u} + n _{d}$ becomes again an even number) in this case with the period corresponding to the flux quantum inside each loop. In the experiments we indeed observe such digital type oscillations of voltage, Fig. 3. The observed periodicity leaves no doubt that the voltage jumps up and down are a consequence of  the change of $n _{u}$ and $n _{d}$. The period of the digital type oscillations in magnetic field $B _{0}  =  \Phi _{0}/S$ corresponds to the flux quantum $\Phi _{0} \approx 20.7  \ Oe \ \mu m^{2}$ inside the loop both with square side $a \approx  20 \ \mu m$ and with $a \approx  4 \ \mu m$: the period $B _{0} \approx 0.053 \ Oe $ observed in the first case corresponds to the area $S = \Phi _{0}/B _{0} \approx 390 \ \mu m^{2} \approx a^{2} $ and the period  $B _{0} \approx 1.2 \ Oe $ observed in the second case corresponds to the area $S = \Phi _{0}/B _{0} \approx 17 \ \mu m^{2} \approx a^{2}$. We observed more than 110 voltage jump up and down at measurement of the structure with square side $a \approx  20 \ \mu m$ in the magnetic field from $B = -3 \ Oe$ to $B = 3 \ Oe$.  

 \begin{figure}[]
\includegraphics{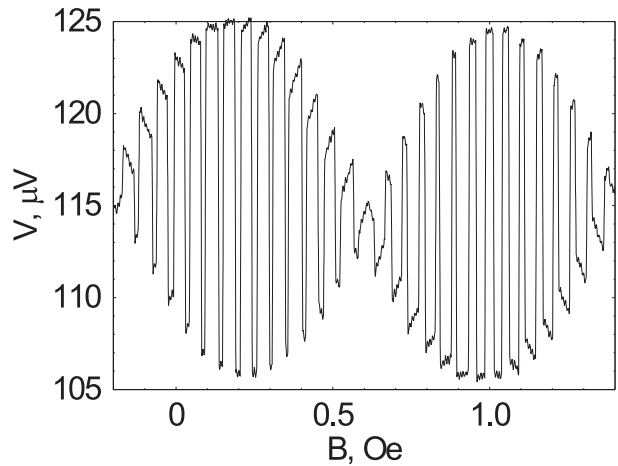}
\caption{\label{fig:epsart} Voltage jumps due to changes of quantum numbers in magnetic field at the bias currents through the structure $I \approx 20 \ nA $ observed at the temperature $T \approx 1.1 \ K $ at measurement of the structures with square side of the loops $a \approx  20 \ \mu m$. A part of magnetic dependence measured from $B = 3 \ Oe$ to $B = -3 \ Oe$ is shown. The voltage jumps up and down with the period $B _{0} \approx 0.053 \ Oe $ and the amplitude of the jumps is modulated with the period $B _{m} \approx 0.8 \ Oe $ in the whole region from $B = 3 \ Oe$ to $B = -3 \ Oe$, see Fig.8 of Supporting Information.}
\end{figure} 

The observation of these jumps has the critical importance for the opportunity to use the DDCI for high sensitive detection of magnetic flux. The conventional theory \cite{Tink75} and numerous experiments \cite{PCJETP07,Tanaka04,APL2016,JETP07J,PCScien07,PLA2012} testify to the sharpness of the change of the probability $P(n)$ of the $n$ state of superconducting loop with enough big value of the persistent current $I _{p,A}$. The theory predicts the jump of the critical current with the $n$ change at $\Phi = (n+0.5) \Phi _{0}$ not only of the double contour interferometer but also, for example, of a superconducting ring with asymmetric link-up of current leads. A simple magnetometer based on the latter prediction was proposed in \cite{Letter2014}. But measurements of aluminium ring with asymmetric link-up of current leads have revealed that a smooth change of its critical current is observed at $\Phi \approx (n+0.5) \Phi _{0}$ instead of the jump which must be observed due to the change of the quantum number from $n$ to $n+1$ \cite{PLA2017}. Our observations of the voltage jumps up to $V _{max} - V _{min} \approx 20 \ \mu V$, Fig.3, and higher mean that the derivative $(\partial \overline{V}/\partial \Phi _{e}) _{I}$ can reach high values when the voltage $\overline{V}$ average in the time $\Theta \gg 1/f$ changes from $\overline{V} \approx V _{min}$ to $\overline{V} \approx V_{max}$ in the narrow interval $\Delta \Phi \approx \Phi _{0}/\epsilon \ll \Phi _{0}/2$: for example $(\partial \overline{V}/\partial \Phi _{e}) _{I}  \approx (V _{max} - V _{min})/\Delta \Phi \approx  20 \ mV/\Phi _{0}$ at  $V _{max} - V _{min} \approx 20 \ \mu V$ and the real value $\epsilon = I _{p,A}\Phi _{0}/k _{B}T \approx 1000$. 

\begin{figure}[]
\includegraphics{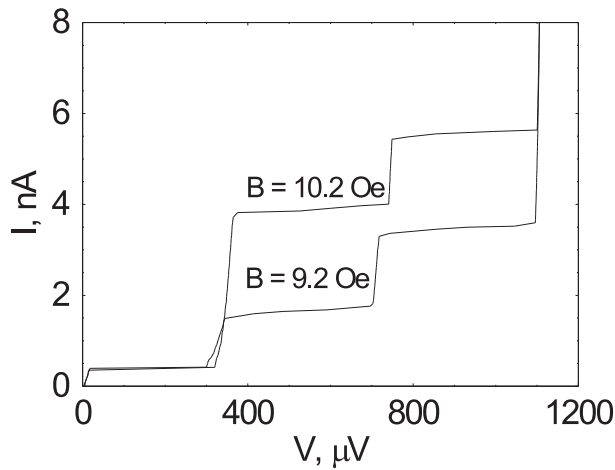} 
\caption{\label{fig:epsart} Current-voltage characteristics of the DDCI structures with square side of the loops $a \approx  4 \ \mu m$ measured at the temperature $T \approx 0.44 \ K $ and different magnetic field $B \approx 9.2 \ Oe$ and $B \approx 10.2 \ Oe$. Two extra steps correspond to the two additional Josephson junctions of the real DDCI at $ l _{u}$ and $r _{d}$.}
\end{figure} 

The experimental corroboration of the voltage jumps, Fig.3, is the central objective of our measurements of the real double contour interferometer shown on Fig.2. Other peculiarities observed at these measurements are considered partly in the Supporting Information. Here we say only about the periodical modulation of the jump amplitude, Fig.3, and some peculiarities of the current-voltage characteristics of the real structures. We assume that the modulation of the jump $V _{max} - V _{min}$ with the period of $B _{m} \approx 0.8 \ Oe $  may be connected with the shift $\approx 0.5 \ \mu m $ of the loops relative to one another in the measured structure, Fig.2, see the Supporting Information. Three steps at the current-voltage characteristics shown in Fig. 4 are observed because of the necessary Josephson junctions $J _{a}$, $J _{b}$ and two unnecessary Josephson junctions $J _{l}$, $J _{r}$ created because of our technology. Two current-voltage characteristics, shown in Fig. 4, measured at different value of magnetic field correspond to even and odd sum $n _{u} + n _{d}$.      

In summary we propose a new type of magnetometer, the DDCI, and present experimental evidence of its opportunity. The DDCI can reach sensitivities better than $20 \ mV/\Phi _{0}$, which exceeds that of dc SQUID by more than one order of magnitude. The effect is due to the strong discreteness of the energy spectrum of the continuous superconducting loop. The advantage of DDCI lies also in circuitry, as one does not need to couple the device to a flux transformer for the measurement of small magnetic fields. The flux transformer is used for the measurement of tiny magnetic fields $B = \Phi /S$ since the area $S$ of the dc SQUID cannot be too large because of the strong screening $\Phi _{I} = LI _{c,j}> \Phi _{0}/2$ and $\beta = 2L I _{c,j}/\Phi _{0} > 1$ in the loop with a high magnetic inductance $L \approx \mu _{0}l$  \cite{SQUID1977}. In the DDCI the magnetic flux induced by the persistent current does not depend on the loop size $l$ since $\Phi _{I} = LI _{p} = (L/L _{k})(n\Phi_{0} - \Phi) \approx (s/\lambda _{L}^{2}(T))(n\Phi_{0} - \Phi)$ \cite{QuSMF2016}. 

 This work has been supported by the Russian Science Foundation, Grant No. 16-12-00070.

\section{\bf Supporting Information }

\subsection{Strong discreteness of the energy spectrum of continuous superconducting loop}
Niels Bohr postulated as far back as 1913 a strong discreteness of the energy spectrum of atoms. According to Bohr's condition $pr = n\hbar $, that the angular momentum $pr$ is an integer multiple of Planck's constant $\hbar = h/2\pi $, the energy discreteness $E_{n+1} - E_{n} = p_{n+1}^{2}/2m - p_{n}^{2}/2m = (2n+1)\hbar ^{2}/2mr^{2}$ decreases with the increase of the orbit radius $r$. The energy spectrum of atom is strongly discrete due to small radius of electron orbits: the energy difference between adjacent permitted states $\Delta E \approx \hbar ^{2}/2mr^{2} \approx 2 \ 10^{-18} \ J$ for the Bohr radius $r_{B} \approx 0.05 \ nm = 5 \ 10^{-11} \ m$ corresponds to the temperature $T = \Delta E/k _{B} \approx 100000 \ K$. The discreteness decreases down to  $\Delta E \approx \hbar ^{2}/2mr^{2} \approx 2 \ 10^{-18} \ J$ corresponding to $T = \Delta E/k _{B} \approx 0.001 \ K$ in a mesoscopic ring with a radius $r \approx 500 \ nm = 5 \ 10^{-7} \ m$. Therefore the quantum phenomena connected with the discrete spectrum, such as the persistent current of electrons, can be observed in nano-rings with a real radius $r > 300 \ nm$ only at very low temperature \cite{PCScien09,PCPRL09}. The energy difference of superconducting loop $\Delta E \approx N _{s}\hbar ^{2}/2mr^{2} $ is much larger due to the impossibility for all $N _{s}$ Cooper pairs in the loop to change their quantum state $n$ individually \cite{FPP2009}. This impossibility of individual motion of quantum particle was postulated first by Lev Landau as far back as 1941 \cite{Landau41} for the description of superfluidity of $^{4}He$ liquid. 

All $N _{s}$ pairs in the loop with the volume $V = ls$, the length $l$ and the section area $s$ are described with the wave function $\Psi = |\Psi |\exp{i\varphi }$, according to the Ginzburg-Landau theory \cite{GL1950}: $|\Psi |^{2} = n_{s}$ is the density of Cooper pairs and $\int_{V}dV |\Psi |^{2} = \oint_{l}dl s|\Psi |^{2} = N_{s} \gg 1$ is the total number of Cooper pairs in the loop. This number  exceeds 100000 in a typical superconducting loop. Therefore its spectrum of the permitted states is strongly discrete: the energy difference between adjacent permitted states of a ring with a radius $r \approx 500 \ nm $ exceeds the value $\Delta E \approx N _{s}\hbar ^{2}/2mr^{2} \approx 2 \ 10^{-21} \ J$ corresponding to the temperature $T = \Delta E/k _{B} \approx 100 \ K$ at $N _{s} > 10^{5}$. The discreteness increases with the increase of all three sizes of the ring $\Delta E \approx N _{s}\hbar ^{2}/2mr^{2} \approx  n_{s}s2\pi r(\hbar ^{2}/2mr^{2}) \propto  (s/r)$ due to the increase of the number of Cooper pairs $N _{s}$. 

The quantization of angular momentum postulated by Bohr may be deduced from the requirement that the complex wave function must be single-valued in any point $l$ of the loop $\Psi = |\Psi |e^{i\varphi } =  |\Psi |e^{i(\varphi + n2\pi )} $. The relation 
$$\oint_{l}dl \bigtriangledown \varphi = 2\pi n \eqno{(1)}$$ 
must be valid for any contour $l$ along which the wave function $\Psi = |\Psi |e^{i\varphi }$ is defined, according to this requirement. The angular momentum of each Cooper pair has a discrete value $n\hbar $ and the total angular momentum of all $N _{s}$ pairs equals $M _{p} = \oint _{l} dl sr\Psi ^{*}\hat{p} \Psi = \oint _{l} dl sr\Psi ^{*} (-i\hbar \nabla )\Psi = sr|\Psi |^{2}\hbar \oint _{l} dl \nabla \varphi  =  N _{s}\hbar n $.

\begin{figure}[] 
\includegraphics{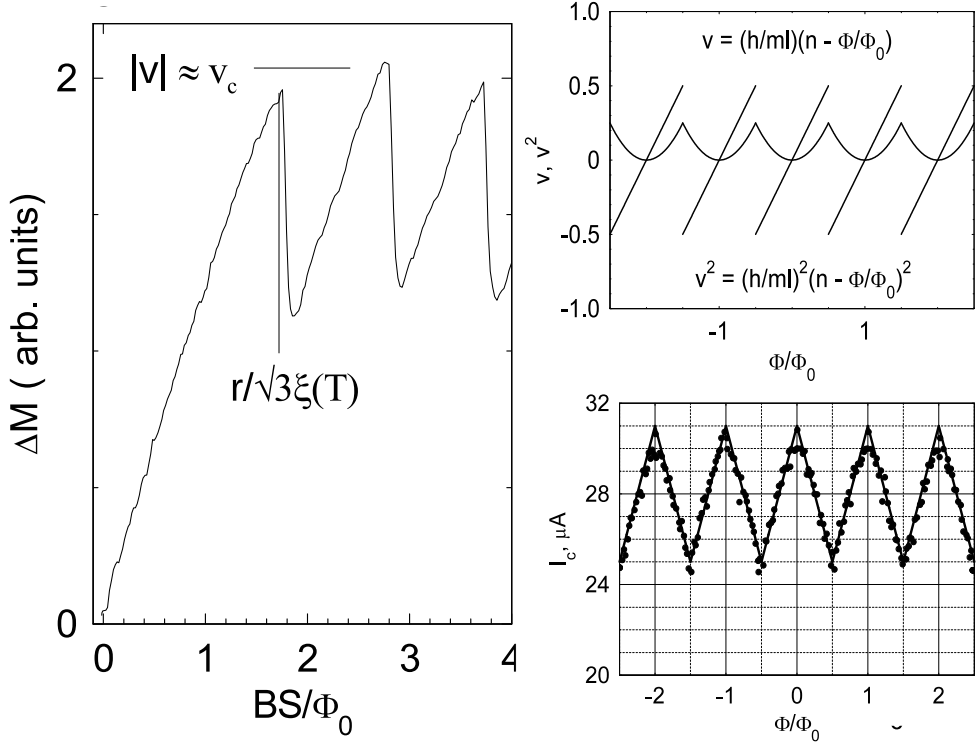}
\caption{\label{fig:epsart} The picture on the left: The virgin magnetization $\Delta M \propto L_{f}I _{p}$ of aluminum ring with the radius $r \approx 1 \ \mu m$ increases up to $|v| \approx  v _{c} $ with the magnetic flux $\Phi \approx BS$. The quantum number changes from $n = 0$ to $n = 1$  $|\Phi /\Phi _{0}| \approx 1.7 \approx  r/\surd 3 \xi (T)$, $n = 1$ to $n = 2$  $|\Phi /\Phi _{0}| \approx 1.7 + 1 = 2.7$, $n = 2$ to $n = 3$  $|\Phi /\Phi _{0}| \approx 1.7 + 2 = 3.7$. The drawing at the upper right: The velocity of Cooper pairs  changes by jump at a definite magnetic flux $\Phi = (n+0.5) \Phi _{0}$ due to the switching of ring segment in the normal state for a while by an external current, a noise or  thermal fluctuations, or due to quantum fluctuations. The two states $n$ and $n+1$ have the opposite velocity $v$ and equal kinetic energy $E _{k} \propto v^{2}$. The picture below on the right: Measurements of the critical current of symmetric aluminum ring with the radius $r \approx 2 \ \mu m$ \cite{JETP07J} corroborate the predominant probability of the $n$ state in the interval $(n-0.5) \Phi _{0} < \Phi < (n+0.5) \Phi _{0}$.}
\end{figure}

According to the canonical definition the gradient operator $\hat{p} = -i\hbar \nabla $  corresponds to the canonical momentum $p = mv + qA$ of a particle with a mass $m$ and a charge $q$ both with $A \neq 0$ and without $A = 0$ magnetic field \cite{LandauL}. The velocity operator $\hat{v} = (\hat{p} - qA)/m = (-i\hbar \nabla - qA)/m$ \cite{FeynmanL}, in contrast to the momentum operator, depends on the magnetic vector potential $A$. Therefore the velocity of Cooper pair with $q = 2e$ 
$$\oint_{l}dlv  = \frac{2\pi \hbar }{m}(n-\frac{\Phi }{\Phi_{0}}) \eqno{(2)}$$
cannot be equal zero when the magnetic flux $\Phi = \oint_{l}dl A$ inside the loop $l$ is not divisible by the flux quantum $\Phi _{0} = 2\pi \hbar /q = \pi \hbar /e \approx 20.7 \ Oe \ \mu m^{2}$. The velocity $ldv/dt = -(q/m)d\Phi /dt$ and the persistent current  $L _{k}dI _{p}/dt = L _{k}sqn_{s}dv/dt = -d\Phi /dt $ change with the magnetic flux $\Phi $ in accordance with the Newton's second law $mdv/dt = qE$, where $E = -\nabla  V - dA/dt$ is the electric field, $L _{k} = ml/ q^{2}n_{s}s = (\lambda _{L}^{2}/s) \mu _{0}l$ is the kinetic inductance of the loop with the length $l$, the cross-sectional area $s = wd$ and the density of Cooper pairs $n_{s}$ and $\lambda _{L} = (m/\mu _{0}q^{2}n_{s})^{0.5} $ is the London penetration depth. The magnetic inductance $L_{f}$ of superconducting loop with a small cross-sectional area $s \ll \lambda _{L}^{2}$  is much lower than the kinetic inductance $L_{f}  \approx \mu _{0}l \ll L _{k} = (\lambda _{L}^{2}/s) \mu _{0}l $ at $s \ll \lambda _{L}^{2}$. The magnetic flux $\Delta \Psi _{I} = L_{f}I _{p}$  induced by the persistent current $I _{p}$ and the energy of the magnetic field $L_{f}I _{p}^{2}/2 \ll L_{k}I _{p}^{2}/2$ are small in this case of weak screening. 

\begin{figure}[]
\includegraphics{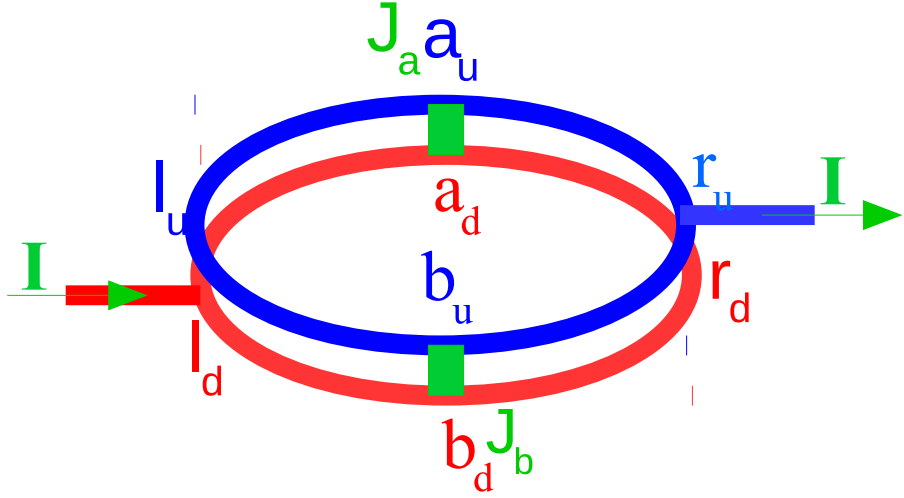}  
\caption{\label{fig:epsart} The scheme of an ideal DDCI. The loops located one above the other and therefore the boundaries of the Josephson junctions $J _{a}$ and $J _{b}$ coincide with the points $a _{u}$, $a _{d}$ and $b _{u}$, $b _{d}$. The critical current of such DDCI should not depend directly on the magnetic flux inside the loops.}  
\end{figure}

The pair velocity may increase up to the depairing velocity $|v| \leq v _{c} = \hbar /m \surd 3 \xi (T)$ in narrow loop segments \cite{Tink75} without a change of the quantum number $n$, Fig.5. The quantum number can change when a density of Cooper pairs diminishes for a while from $n_{s} = 2n_{s0}/3$ \cite{Tink75} to $n_{s} = 0$ at $|v| \geq  v _{c} $  in a segment, where $n_{s0}$ is the density at $|v| = 0$. The quantum number of homogeneous ring changes at $|n - \Phi /\Phi _{0}| \approx r/\surd 3 \xi (T)$, Fig.5,  when the velocity increases only because of magnetic flux $\Phi \approx BS$ change in accordance with (2). The jump of the velocity (2) and the persistent current, observed in this case \cite{Geim2003,Moler2007,Nat2016}, cannot be used for a magnetometer designing because of the problem with a hysteretic field characteristic. The quantum number can change from $n$ to $n+1$ at a definite magnetic flux $\Phi = (n+0.5) \Phi _{0}$, Fig.5, due to the switching of a segment in the normal state for a while by an external current at $T < T _{c}$ \cite{PRB2014C}, by thermal fluctuations at $T \approx T _{c}$ \cite{JLTP1998} or quantum tunneling \cite{Tanaka2002}.

\begin{figure}[]
\includegraphics{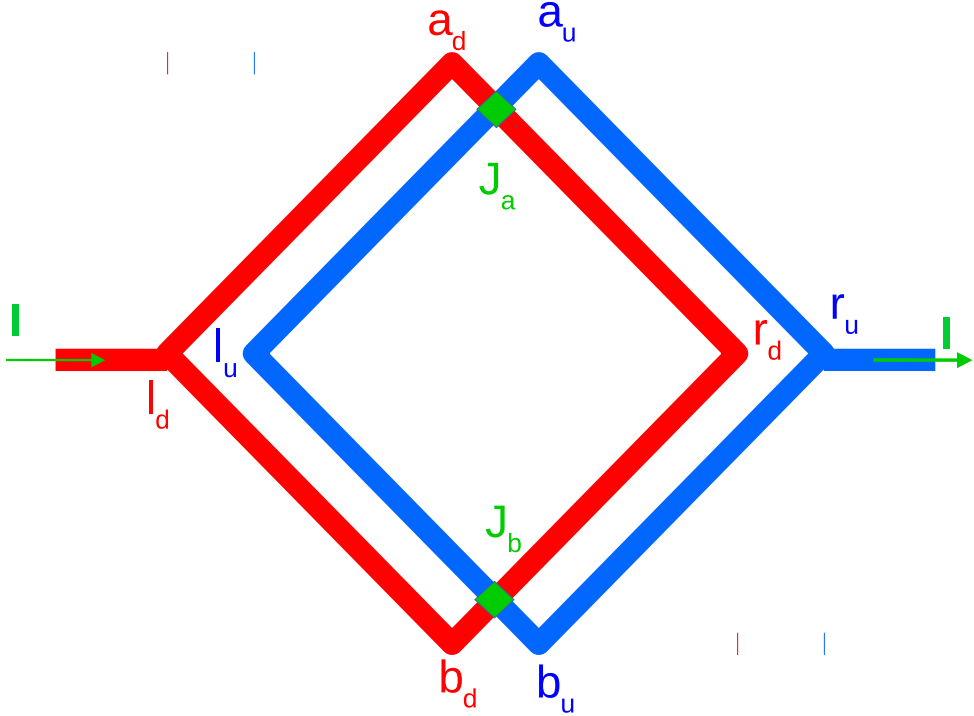}
\caption{\label{fig:epsart} The scheme of the DDCI with the loops shifted relative to each other. The boundaries of the Josephson junctions $J _{a}$ and $J _{b}$ do not coincide with the points $a _{u}$, $a _{d}$ and $b _{u}$, $b _{d}$. Therefore the critical current may depend directly on the magnetic flux inside the loops.}
\end{figure} 

The ring is switched in the normal state for a while by the external current in the process of measuring the critical current \cite{PRB2014C}. The measurements \cite{PCJETP07} corroborate that the quantum number of the ring corresponds to the minimal kinetic energy $\propto v^{2}$ with the predominant probability after its coming back in superconducting state. The oscillations of the critical current of a symmetric superconducting ring, Fig.5, observed for example in \cite{JETP07J}, are similar to the one of a conventional dc SQUID, i.e. superconducting loop with two Josephson junctions  \cite{Barone}. But there is a principal difference between superconducting loops with and without the Josephson junction. The phase change at a complete turn along the loop with  Josephson junctions depends on the sum of the phase differences on the Josephson junctions and the magnetic field inside the loop: $2\pi n = \sum _{i}\varphi _{i} - 2\pi \Phi /\Phi _{0}$ \cite{Barone}. The persistent current in the loop with one Josephson junction, i.e. in an rf SQUID, $I _{p} = I _{c,j}\sin \varphi =  I _{c,j}\sin (2\pi n + 2\pi \Phi /\Phi _{0}) =  I _{c,j}\sin (2\pi \Phi /\Phi _{0})$ should not depend on the quantum number $n$ because of the mathematical equality $\sin (\varphi + 2\pi n) \equiv \sin (\varphi )$. Therefore no jump connected with the $n$ change should be observed in this case. The jump of the persistent current should be observed in the loop without the Josephson junction. This jump should not result to the jump of the critical current of a symmetric ring. The jump of the critical current must be observed at measurements of asymmetric rings. But this jump is not observed \cite{PCJETP07,PLA2017} for some strange reason.

\subsection{Superconducting current through the differential double contour interferometer}
The phases $\varphi $ of the wave functions $\Psi = |\Psi |\exp{i\varphi }$ describing superconducting state of the two loops of the differential double contour interferometer are connected due to the two Josephson junctions $J _{a}$ and $J _{b}$, Fig.6. According to the Josephson relation $I _{s} = I _{c,j}\sin \varphi $ the superconducting currents depend of the phase difference $\varphi _{a}$ and $\varphi _{b}$ between the boundaries of the Josephson junctions $J _{a}$ $\varphi _{a} = \varphi _{a,d} - \varphi _{a,u}$ and $J _{b}$ $\varphi _{b} = \varphi _{b,d} - \varphi _{b,u}$, where $\varphi _{a,d}$, $\varphi _{a,u}$,  $\varphi _{b,d}$ and $\varphi _{b,u}$ are the phase of the wave functions at the points $a _{d}$, $a _{u}$,  $b _{d}$ and $b _{u}$. The total superconducting current through the two Josephson junctions   
$$I _{s} = I _{a}\sin  \varphi _{a} + I _{b}\sin  \varphi _{b} \eqno{(3)}$$ 
should depend on the phases $\varphi _{a,d}$, $\varphi _{a,u}$,  $\varphi _{b,d}$ and $\varphi _{b,u}$. The phase change from $a _{d}$ to $b _{d}$ and from $b _{d}$ to $a _{d}$, clockwise for example, should be equal $2\pi n _{d}$ and the phase change from $a _{u}$ to $b _{u}$ and from $b _{u}$ to $a _{u}$  should be equal $2\pi n _{u}$ according to (1). The phase changes from $a _{d}$ to $b _{d}$ and from $b _{d}$ to $a _{d}$ are equal when the velocity of Cooper pairs in the ring halves is the same. The velocity should be the same in a homogeneous loop with the persistent current exceeding strongly the superconducting current (3) through the two Josephson junctions $I _{p} \gg I _{s}$. In this case the phase differences of the Josephson junctions $J _{a}$ and $J _{b}$ are connected with the quantum numbers of the loops by the simple relation $\varphi _{a} - \varphi _{b} = (\varphi _{a,d} - \varphi _{a,u}) - (\varphi _{b,d} - \varphi _{b,u}) =  \pi (n _{u} + n _{d})$ due to the equality $\varphi _{a,d} - \varphi _{b,d} =  \pi n _{d}$ and $\varphi _{b,d} - \varphi _{a,d} =  \pi n _{d}$. According to this relation $\varphi _{b} = \varphi _{a} - \pi (n _{u} + n _{d})$ and the superconducting current (3) depends on the single phase difference $\varphi _{a}$ and two quantum numbers $I _{s} = (I _{a} + I _{b})\sin  \varphi _{a} = 2I _{cj}\sin  \varphi _{a}$ at $I _{a}= I _{b} = I _{cj}$ when the sum $n _{u} + n _{d}$ is even and $I _{s} = (I _{a} - I _{b})\sin  \varphi _{a} = 0$ at $I _{a}= I _{b}$ when $n _{u} + n _{d}$ is odd. 

\begin{figure}[]
\includegraphics{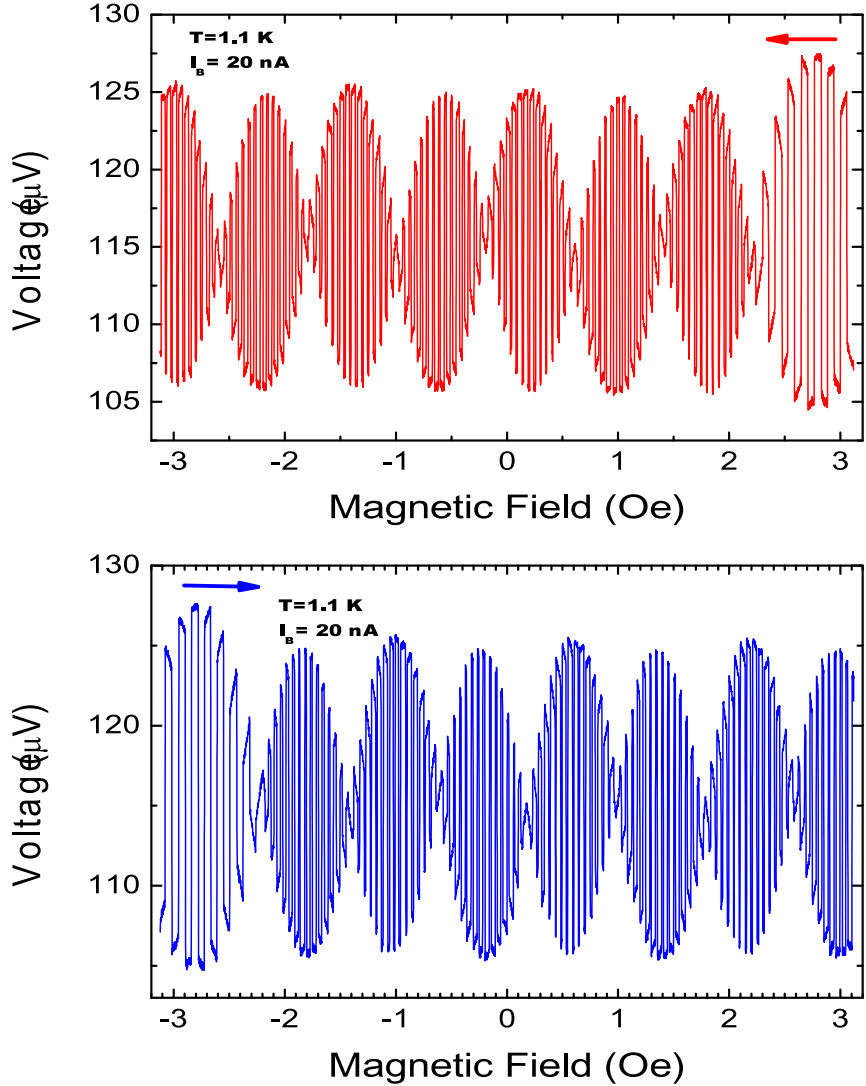} 
\caption{\label{fig:epsart} Voltage oscillations in the DDCI with the side of the loops $a  \approx 20 \ \mu m$ measured in the opposite directions of magnetic field sweep, from $B \approx +3 \ Oe$ to $ B \approx -3 \ Oe$ and $B \approx -3 \ Oe$ to $ B \approx +3 \ Oe$.}
\end{figure} 

\subsection{The differential double contour interferometer with shifted loops}  
The phase differences between the boundaries of the Josephson junctions are connected with the quantum numbers of superconducting loops by the simple relation $\varphi _{a} - \varphi _{b} = \pi (n _{u} - n _{d})$ due to the overlap of the points $a _{d}$, $J _{a}$, $a _{u}$ and $b _{d}$, $J _{b}$, $b _{u}$ of the ideal structure shown on Fig.6. These points are shifted with respect to each other, Fig.7, in the real structure because of  the simple shadow evaporation technique used in our work for the DDCI fabrication. The upper loop is shifted relatively the bottom one on $a _{sh} \approx 0.6 \ \mu m$ because of this technique. The path from the upper (bottom) boundary of the Josephson junction $J _{a}$ to the upper (bottom) boundary of the $J _{b}$ through $l _{u}$ ($r _{d}$) is not equal the path through $r _{u}$ ($l _{d}$), Fig.7. The first path is longer than the second one on $4a _{sh}\surd{2}$. The path along the contour $l _{d} - a _{d} - J _{a} - a _{u} - r _{u} - b _{u} - J _{b} - b _{d} - l _{d}$ increases on the value $4a _{di} = 4a _{sh}\surd 2$, Fig.7. Therefore the superconducting current through the real DDCI used in our work for the experimental investigations 
$$I _{s} = I _{a}\sin  \varphi _{a} + I _{b}\sin  (\varphi _{a}+ \pi (1+\frac{a _{di}}{a})(n _{u} + n _{d}) + 2\pi \frac{a _{di}aB}{\Phi _{0}}) \eqno{(4)}$$
depends not only on the phase difference $\varphi _{a}$ and the quantum numbers $n _{u} - n _{d}$ but also on the magnetic field value $B$. The term $2\pi a _{di}aB/\Phi _{0}$ in the relation can explain the modulation of the jump amplitude with the period $B _{m} \approx 0.8 \ Oe$ observed at measurements of the DDCI with the side of the loops $a  \approx 20 \ \mu m$, Fig.8. 

\subsection{What is the maximal magnetic field in which the voltage jumps may be observed?} 
The jumps of the critical current of the DDCI and the voltage jumps at a biased current can be observed until their loops are in superconducting state. Magnetic field depresses superconductivity in a strip with a finite width $w$. The value of the depression depends of the temperature and the stripe width. For example the critical current of the aluminum strip with $w \approx 0.6 \ \mu m$ decreases in two time at the magnetic field $B \approx  20 \ Oe$ and in four time at $B \approx 30 \ Oe$ at the temperature $T \approx 0.986T _{c}$, see Fig.9 of \cite{PCJETP07}. More than 1000 jumps may be observed in the interval $30 \ Oe < B < 30 \ Oe$ at the measurement of the DDCI with the side of the loops $a  \approx 20 \ \mu m$ with the period of jumps $B _{0} \approx 0.053 \ Oe$. We made the measurement in the interval $3 \ Oe < B < 3 \ Oe$ and observed more than 100 jumps, Fig.8. The depression of superconductivity by magnetic field decreases with the temperature $T$ and width $w$ decrease. The voltage jumps may be observed in the interval of magnetic field much wider than $30 \ Oe < B < 30 \ Oe$ when $T < 0.986T _{c}$ and $w < 0.6 \ \mu m$. The number of the jumps may be much more than 1000.  

We demonstrate on Fig.8 the hysteretic field characteristic of our DDCI. The magnetic field at which the quantum numbers of the loops change depends on the direction of sweep because the loop is switched in normal state for a while by the persistent current $I _{p}$ at $|n - \Phi /\Phi _{0}| \approx r/\surd 3 \xi (T)$ rather than by an external current $I _{ext}$, a noise or thermal fluctuations. The quantum number of the superconducting loop can change both on 1 and 2, 3… in this case \cite{Geim2003,Moler2007}. It is observed on Fig.8 that the period of the voltage jumps $B _{0} \approx 0.11 \ Oe$ in the beginning of the sweep and $B _{0} \approx 0.053 \ Oe$ in the continuation of the sweep. The period $B _{0} \approx 0.053 \ Oe$ corresponds to the flux quantum $B _{0}S = \Phi _{0}  \approx 20.7 \ Oe \ \mu m^{2}$  in the loop with the area $S = \Phi _{0}/B _{0} \approx 390 \ \mu m^{2} \approx (19.8 \ \mu m)^{2}$. Consequently, the  observations shown on Fig.8 mean that the quantum numbers of the DDCI loops change on 2 in the beginning of the sweep and 1 in the continuation of the sweep. It is not clear why this difference is observed.


\end{document}